\documentclass[showpacs,preprintnumbers,pra,twocolumn,amsmath,amssymb,amsfonts,amsthm]{revtex4}
\usepackage{amssymb}
\usepackage{amsbsy}
\usepackage{epsf}

\usepackage[cp866nav]{inputenc}
\usepackage{revsymb}
\usepackage[T2A,OT1]{fontenc}
\usepackage{graphicx}

\newcommand{\be}{\begin{equation}}
\newcommand{\ee}{\end{equation}}
\newcommand{\bea}{\begin{eqnarray}}
\newcommand{\eea}{\end{eqnarray}}

\begin{document}

\title
{Peculiarities of qubit initial-state preparation by nonselective
measurements on an overcomplete basis}
\author{V.V. Ignatyuk}
\affiliation{Institute for Condensed Matter Physics, 1
Svientsitskii Street, 79011, Lviv, Ukraine}

\date{\today}

\begin{abstract}
We consider the qubit initial state preparation due to the
nonselective measurements on an overcomplete basis, when the
number of outcomes $N=3$. To be specific, we have chosen the
dephasing model and applied the conditions for a post-measurement
state to ensure a gradual coherence enhancement at the initial
stage of the system evolution. It is shown that contrary to the
von Neumann-L\"uders projection scheme with $N=2$, the most mixed
post-measurement state of the qubit in the general case can depend
on the bath temperature. This remarkable feature allows one to
consider a ``temperature controlled'' initial purification of the
open quantum system. We also consider some kinds of the repeated
nonselective measurements from the viewpoint of an eventual system
purification and analyze the important peculiarities of such
schemes.

\pacs{03.65.Ta, 03.65.Yz, 03.67.Pp}
\end{abstract}

\maketitle

\section{Introduction\label{Introduct}}
The initial state preparation is an essential point in the
dynamics of quantum open systems . Due to unavoidable
interactions, there generally exist initial correlations between
the system and its environment. Since any physical process of
preparation of the initial state of the system will affect the
state of the environment, a natural question appears: how the
preparation procedure and initial correlations act upon subsequent
evolution of the system. Of special interest is the decoherence
phenomenon (the environmentally induced destruction of quantum
coherence), which plays a crucial role in the dynamics of
two-state systems (qubits), which are widely recognized as the
elementary carriers of quantum
information~\cite{q-measurement1,39min}. Different aspects of
influence of the initial state preparation on the system dynamics
were discussed by  many authors (see, e.g.,
Refs.~\cite{PRA2010,PRA2012,Luczka} and references therein).
Usually \cite{MMR2012,PRA2013,PRA2013-2}, the initial correlations
lead to the onset of the additional dissipation channel, since the
system evolves from the \textit{post-measurement} state, which
differs from the equilibrium (Gibbsian) one. Nevertheless, the
opposite picture can be observed too \cite{ourPRA}, and there is a
coherence enhancement in the system ``qubit--thermal bath'',
prepared initially by the nonselective measurement
\cite{BP-Book,Krauss} or even by some more specific preparation
procedure ~\cite{Luczka}.

The above mentioned findings can be helpful from the viewpoint of
quantum measurements and control \cite{measur}, since the time
evolution of the qubit (or the chains of qubits) strongly depends
on the initial state of the system. Though the thermal and vacuum
fluctuations of the environment are widely believed
\cite{BP-Book,q-measurement} to eliminate the influence of the
initial correlations in the system, it cannot be the case if a
residence time of the quantum register before its switching is
less (or comparable) with the dephasing time
\cite{MMR2012,ourPRA}. In this context, the most desirable case is
if the system can even be purified dynamically
\cite{ourPRA,Luczka} by interaction with the environment to
protect an eventual loss of information stored in it.

On the other hand, it is interesting to investigate the initial
system preparation not only from the viewpoint of the subsequent
qubit dynamics. Though these problems were studied usually from
pure mathematical points of view \cite{Holevo2001,SRM}, novel
trends and achievements in the quantum measurements and control
\cite{measur,q-control,control-PRL} dictate new challenges for
physicists concerning special \textit{post-measurement} system
preparation. In particular, some questions that could be imposed,
sound as follows: i) what factors influence the system purity (or
mixedness); ii) what ``risks'' should be avoided when manipulating
the open quantum system at the initial stage of its preparation;
iii) how to separate the factors contributing to the system
purification from those leading to undesirable mixed\-ness.

In this paper, we make an attempt to give some particular answers
to these questions. To be specific, we choose the spin--boson
system described by the dephasing model
\cite{Luczka1,MMR2012,ourPRA}. We consider a special kind of the
nonselective measurements \cite{BP-Book}, where the number $N$ of
outcomes is larger than the rank $n$ of the initial density matrix
(for the qubits $n=2$). Unlike the well-known von Neumann-L\"uders
projection measurements studied earlier \cite{ourPRA}, this more
general scheme is associated with the notion of the
 positive operator-valued measure (POVM)~\cite{Holevo2001,SRM}.
In this scheme, the basis of
 measurement state turns out to be \textit{overcomplete}, and the measurement
 state vectors can no longer be orthonormal.
 In fact, we deal
 with a non-orthogonal resolution of the identity \cite{resol_of_1} with
all specific features as compared to the orthogonal one.

To restrict ourselves by a particular choice, we impose the
\textit{post-measurement} conditions considered in
Ref.~\cite{ourPRA}, which ensure a gradual coherence enhancement
at the initial stage of the system evolution. Of a particular
interest is the system purity and its response to the change of
both internal (e.g., temperature fluctuations) and external
(measurement state vectors) factors. We show that the most mixed
state (least desirable from the viewpoint of a quantum register
design) can be sensitive to the bath temperature that gives us a
possibility to perform a ``temperature controlled'' initial
purification of the open quantum system. At some other specific
choices of the control parameters, the minimum of the system
purity is insensitive to temperature. Moreover, in some particular
cases this minimum equals 1/2 that corresponds to a completely
decohered system with an equal population of the levels.
Naturally, realization of such a scenario should be avoided when
manipulating the qubit at the moment of its preparation.

We also consider some special kinds of the repeated nonselective
measurements, which have been performed instantly at the initial
time $t=0$. Our purpose is to study at what conditions the purity
can definitely grow after the second measurement, and what
measurement settings are unfavorable from the viewpoint of the
system purification.

  The structure of the paper is as follows. Sec.~\ref{secII} contains a brief review of
the measurement schemes for open quantum systems at an arbitrary
  number $N$ of the outcomes and its application to qubits. We
emphasize the distinctive features of $N=2$ and $N>2$ cases.
  In Sec.~\ref{secIII} we present briefly the main equations
  governing the qubit dynamics in the framework of the dephasing
  model. The condition leading to the gradual coherence
  enhancement at the initial stage of the system evolution is presented in the explicit form.
This condition is taken to be a cornerstone during the study of
the initial system purity in Sec.~\ref{secIV}, when a number of
outcomes at the nonselective measurement is equal to three. We
consider different cases of the \textit{post-measurement} system
preparation, at which the most mixed state of the qubit is
realized. A thorough analysis of a dependence of the about
mentioned state on the internal and external control parameters is
given. A response of the open quantum system to the series of
repeated nonselective measurements performed instantly is studied
in Sec.~\ref{secV}. Finally, we draw the conclusions in
Sec.~\ref{secVI}.

\section{The nonselective preparation measurements and the initial density matrix\label{secII}}

Suppose that at all times $t<0$ an open system $S$ is in thermal
equilibrium with a heat bath $B$, and at time zero one makes a
measurement on the system $S$ only. According to the general
principles of quantum measurement theory
\cite{BP-Book,q-measurement,Krauss}, the state of the composite
system ($S+B $) after the measurement is characterized by the
density matrix
 \begin{equation}
    \label{rho-SB-0}
 \varrho^{}_{SB}(0)= \sum_{m=1}^N\Omega^{}_{m}\varrho^{}_{\text{eq}}\Omega^{\dagger}_{m},
 \end{equation}
where $\varrho^{}_{\text{eq}}$ is the equilibrium density matrix
at temperature $T$. Linear operators $\Omega^{}_{m}$ act in the
Hilbert space ${\cal H}$ of the system $S$ and correspond to
 possible outcomes $m=\{1,\ldots,N\}$ of the measurement.
In a particular case of a
 \textit{selective measurement\/}, the system $S$ is prepared in some
pure state $|\psi\rangle $. Then the sum in Eq. (\ref{rho-SB-0})
collapses into a single term, so that
  \begin{equation}
    \label{rho-SB-selec}
     \varrho^{}_{SB}(0)=\frac{1}{Z}\, P^{}_{\psi}\varrho^{}_{\text{eq}}P^{}_{\psi},
  \end{equation}
where $ P^{}_{\psi}=|\psi\rangle \langle\psi| $ is the projector
onto a pure quantum state $|\psi\rangle $ and $Z$ is the
normalization factor. In general, the density matrix
(\ref{rho-SB-0}) describes the resulting ensemble after a
\textit{nonselective measurement\/}, in which the outcome $m$  may
be viewed as a classical random number with the probability
distribution
    \begin{equation}
   \label{p-m}
  w(m)=\text{Tr}\left\{F^{}_{m}\varrho^{}_{\rm eq}\right\}.
  \end{equation}
Positive operators
  $F^{}_{m}=\Omega^{\dagger}_{m}\Omega^{}_{m}$ are called the ``effects''.
   Here and in what follows, $\text{Tr}$ denotes the trace over the Hilbert space of the
    composite ($S+B $) system, while the symbols $\text{Tr}^{}_{S}$ and $\text{Tr}^{}_{B}$
     will be used to denote the partial traces over the Hilbert spaces of the system
     $S$ and the heat bath, respectively.
For $w(m)$  to be normalized to 1, the effects $F^{}_{m}$ must
satisfy the
 normalization condition (the resolution of the identity \cite{resol_of_1})
    \begin{equation}
    \label{Fm-NormCond}
  \sum_{m=1}^N F^{}_{m}\equiv \sum_{m=1}^N\Omega^{\dagger}_{m}\Omega^{}_{m}=I,
  \end{equation}
where $I$ is the unit operator.

 The precise form of $\Omega^{}_{m}$ is determined by the details of the measuring device
 \cite{ourPRA,BP-Book} and could be applied both to \textit{approximate\/} measurements,  where the spectrum of the observed variable
 is measured with a
  finite resolution~\cite{BP-Book,Krauss}, and to the case of an
 infinite resolution. The effects
  \begin{equation}
    \label{Fm-ideal}
   F^{}_{m}= |\psi^{}_{m}\rangle \langle\psi^{}_{m}|,
  \end{equation}
and the accompanying them $\Omega$-operators
  \begin{equation}
     \label{Omega-m-gen}
  \Omega^{}_{m}=|\varphi^{}_{m}\rangle \langle \psi^{}_{m}|
  \end{equation}
  are constructed on the quantum pure
  states -- the measurement state vectors $|\psi_m\rangle$,
  $|\varphi_m\rangle$.

 A few remarks are needed here. In the case when the number of
 outcomes $N$ equals the rank $n$ of the system density matrix
 $\varrho_S(0)$ (for a qubit $n=2$), the states $|\psi^{}_{m}\rangle$ form an
   orthonormal basis. If $N>n$, taking the trace in normalization condition
   (\ref{Fm-NormCond}) over the system degrees of freedom, one obtains
   \bea\label{psi-psi}
\langle\psi_m|\psi_m\rangle=\frac{n}{N}. \eea It means that in
contrast to the von Neumann-L\"uders scheme, the effects $F_m$ are
not projection operators, since $F_m^2\ne F_m$. Moreover, in such
a case one deals with the non-orthogonal measurement state vectors
since the basis $|\psi_m\rangle$ is \textit{overcomplete}
\cite{Holevo2001}, and the scalar product
$\langle\psi_m|\psi_n\rangle$ is not equal to zero at $m\ne n$,
$\{m,n\}=\{1,\ldots,N\}$. Thus Eq.~(\ref{Fm-NormCond}) should be
treated as a non-orthogonal resolution of the identity
\cite{Holevo2001,Krauss,resol_of_1}.

The subset of vectors $|\varphi_m\rangle$ can be chosen to be
normalized, othewise being arbitrary. In Ref.~\cite{ourPRA} we
have shown that in the von Neumann-L\"uders projection
measurements with $N=2$, at some special cases, when the Gram
operator \cite{Holevo2001} \bea\label{DualScheme}
G=\sum\limits_{m=1}^{N}\Omega_m\Omega^{\dagger}_m=\frac{2}{N}
\sum\limits_{m=1}^N|\varphi_m\rangle\langle\varphi_m|, \eea has a
diagonal matrix, $G=diag$, (including the case $G=I$), the
enhancement of the coherence for this special model spin-boson
system is provided at the initial stage of its evolution. We will
touch upon this interesting phenomenon in more details in
Sec.~\ref{secIII}.

 Let us now apply the above general construction to a qubit.
 In the formal ``spin'' representation,
 the canonical orthonormal basis states of the qubit are
  \begin{equation}
    \label{qbit-can-bas}
  |0\rangle=
  \left(
   \begin{array}{c}
   0\\ 1
   \end{array}
  \right),
  \quad
    |1\rangle=
  \left(
   \begin{array}{c}
   1\\ 0
   \end{array}
  \right).
  \end{equation}
 Two subsets of the measurement state vectors can be
 presented as
 \bea\label{psi-phi}\nonumber
&&|\psi_m\rangle=a_0^{(m)}|0\rangle+a_1^{(m)}|1\rangle,\\
&&|\varphi_m\rangle=b_0^{(m)}|0\rangle+b_1^{(m)}|1\rangle,
 \eea
where $a_i^{(m)}$ and $b_i^{(m)}$ ($i=\{0,1\}$,
$m=\{1,\ldots,N\}$) denote the corresponding amplitudes.
Alternatively, the state measurement vectors (\ref{psi-phi}) can
be presented as
    \begin{equation}
      \label{a-qubit}
     |\psi_m\rangle=\sqrt{\frac{2}{N}}
     \left(
     \begin{array}{c}
     \displaystyle {e}^{-i\phi_a^{(m)}/2}\,\cos(\theta_a^{(m)}/2)\,\\[7pt]
       \displaystyle {e}^{i\phi_a^{(m)}/2}\,\sin (\theta_a^{(m)}/2)
      \end{array}
     \right),
    \end{equation}
        \begin{equation}
      \label{b-qubit}
     |\varphi_m\rangle=
     \left(
     \begin{array}{c}
     \displaystyle {e}^{-i\phi_b^{(m)}/2}\,\cos(\theta_b^{(m)}/2)\,\\[7pt]
       \displaystyle {e}^{i\phi_b^{(m)}/2}\,\sin (\theta_b^{(m)}/2)
      \end{array}
     \right),
    \end{equation}
     where the parameters $\phi_a^{(m)}$, $\phi_b^{(m)}$
     and $\theta_a^{(m)}$, $\theta_b^{(m)}$ are the Euler angles of the unit vector, describing the direction of the
     ``spin'' on the Bloch sphere
     \cite{MMR2012,ourPRA,Holevo2001}.
It is easy to express the amplitudes $a_i^{(m)}$, $b_i^{(m)}$ in
Eqs.~(\ref{psi-phi}) via the corresponding Euler angles according
to (\ref{a-qubit})-(\ref{b-qubit}).

Expression (\ref{rho-SB-0}) for the \textit{post-measurement}
density matrix can now be cast in a more transparent form,
\bea\label{rho-SB-01} \varrho_{SB}(0)=\sum\limits_{m=1}^N
\langle\psi_m|\varrho_{eq}|\psi_m\rangle|\varphi_m\rangle\langle\varphi_m|,\eea
From Eq.~(\ref{rho-SB-01}) the reduced density matrix
$\varrho_S(0)$ can be ob\-tain\-ed after taking the trace over
bath variables, \bea\label{rho-S(0)}
\varrho_S(0)=\sum\limits_{m=1}^N\omega_m
|\varphi_m\rangle\langle\varphi_m|,\,\,
\omega_m=\mbox{Tr}_B\langle\psi_m|\varrho_{eq}|\psi_m\rangle|.\eea
It is seen from Eq.~(\ref{rho-S(0)}) that contrary to the
\textit{selective} case (see Eq.~(\ref{rho-SB-selec}) for
comparison), the qubit is now prepared in the mixed state, where
the probabilities $\omega_m$ are determined by the type of
interactions in the composite ($S+B$) system and depend on the
parameters of the qubit and the thermal bath as well as on the
``external'' parameters (the amplitudes $a_i^{(m)}$ or Euler
angles $\phi_a^{(m)}$, $\theta_a^{(m)}$ of the measurement state
vectors $|\psi_m\rangle$).

A very general formula (\ref{rho-S(0)}) will be used in
Sec.~\ref{secIV} to calculate the \textit{post-measurement} purity
of the qubit, described by the dephasing model
\cite{MMR2012,ourPRA}. It will be shown to lead to quite
unexpected consequences as compared to the von Neumann-L\"uders
projection measurement.

\section{The dephasing model: dynamics of decoherence\label{secIII}}

 Our central goal in this section is to present briefly the basic equations
  governing the qubit dynamics in the framework of the dephasing
  model. This quite a simple model is known to describe the main decoherence mechanism
 for certain types of system-environment  interactions~\cite{MMR2012,Luczka1,Unruh,MR-CMP2012,myCMP}.
 In this model, a two-state system (qubit) ($S$) is coupled to the bath ($B$) of
harmonic oscillators. Using the ``spin'' representation for the
qubit with the basic states
 (\ref{qbit-can-bas}), the total Hamiltonian in the Schr\"odinger picture is taken to be
 (in our units $\hbar = 1$)
    \begin{eqnarray}
    \label{H}
    \nonumber
  && H=H_S+H_B+H_{int}\\
   && {}=\frac{\omega_0}{2}\sigma_3
   +\sum\limits_k\omega_k b^{\dagger}_k b_k
   +\sigma_3\sum\limits_k(g_k b^{\dagger}_k+g^*_k b_k),
   \end{eqnarray}
where $\omega_0$ is the energy difference between the excited
$|1\rangle$ and the ground $|0\rangle$ states of the qubit while
$\sigma_3=\left(\begin{array}{cc}1&0\\0&-1\end{array}\right)$
denotes the third of the Pauli matrices. Bosonic creation and
annihilation operators $b^{\dagger}_k$ and $b_k$ correspond to the
$k$th bath mode with the frequency $\omega_k$, and $g_k$ are the
 coupling constants.

  The dephasing model (\ref{H}) has two distinctive features. First,
  the operator $\sigma^{}_{3}$ commutes with the Hamiltonian and,
   consequently, the average populations of the qubit states do not depend on time.
 Thus we have a unique situation where the system relaxation may be interpreted
physically as ``pure'' decoherence and exchange of
entropy~\cite{MR-CMP2012,myCMP},
 rather than dissipation of energy. Second, in this model equations of motions for
  all relevant operators can be solved exactly~\cite{MMR2012}. This allows one to
   study  the time evolution of the coherences for different initial
   conditions. Here we leave out many details for which we refer to
    Ref.~\cite{MMR2012} and simply quote some important results.

 For a qubit, the quantities of principal interest are the
  \textit{coherences\/} $\langle \sigma^{}_{\pm}(t)\rangle\equiv\mbox{Tr}[\sigma^{}_{\pm}(t)\varrho_{SB}(0)]$, where
   $\sigma^{}_{\pm}(t)$ denotes the spin rais\-ing/lowering
   operators in  the Heisenberg picture \cite{MMR2012} (in what follows, the notation $\langle \ldots\rangle$ will be used
   for averages with the \textit{post-measurement} density matrix $\varrho_{SB}(0)$).
  It is easy to verify that the coherences are related directly to
   the off-diagonal elements of the reduced density matrix of the qubit
   \cite{MMR2012,ourPRA}.

    If the initial state is prepared by a nonselective measurement with the density matrix (\ref{rho-SB-0}) of the composite
    system, and taking into account that the time dependence of both spin and bosonic operators in
   the dephasing model can be evaluated exactly
   \cite{MMR2012,Luczka1,ourPRA},
    we get the following relation for the coherences:
       \bea
    \label{coh1}\nonumber
   \langle\sigma_{\pm}(t)\rangle=\frac{1}{\mbox{Tr} e^{-\beta H}}\sum_{m=1}^N\text{Tr}
   \left[\Omega_m^{\dagger}\sigma_{\pm}(t)\Omega_m e^{-\beta H}\right]\\
= \langle\sigma_{\pm}\rangle \exp[\pm i(\omega_0 t+\chi(t)]
\exp[-\widetilde\gamma(t)],
  \eea
  where $\beta=1/k^{}_{\text{B}}T $ and
  \bea
     \label{gamma-eff}
  \widetilde\gamma(t)=\gamma(t)+ \gamma^{}_{cor}(t)
    \eea
denotes the effective decoherence function. The first term in
(\ref{gamma-eff}) describes a loss of the coherence due to the
vacuum and thermal fluctuations in the bath~\cite{MMR2012,ourPRA}.
It is a second-order function in the coupling parameter $g_k$, and
can be expressed via so-called spectral density function in the
continuum limit of the bath modes \cite{MMR2012,ourPRA}. An
appearance of the second term $\gamma_{cor}(t)$ in
(\ref{gamma-eff}) as well as the phase shift $\chi(t)$ in
(\ref{coh1}) is explained by the fact that the density matrix of
the composite system evolves in time starting from the
\textit{post-measurement} value \bea\label{rho-SB-gen}
&&\varrho_{SB}(0)=\frac{1}{Z}\sum\limits_{m=1}^N|\varphi_m\rangle\langle\varphi_m|\\
&&\nonumber\times\left\{ |a_0^{(m)}|^2 e^{\beta\omega_0/2}
e^{-\beta H_B^{(-)}}+|a_1 ^{(m)}|^2 e^{-\beta\omega_0/2} e^{-\beta
H_B^{(+)}}\right\}, \eea which is defined by the
system-environment initial entanglement. In Eq.~(\ref{rho-SB-gen})
  \begin{equation}
  \label{HBpm}
 H_B^{(\pm)}=\sum\limits_k\omega_k b^{\dagger}_k
 b_k\pm\sum\limits_k(g_k b^{\dagger}_k+g^*_k b_k),
 \end{equation}
 \vspace{-2mm}
and \bea\label{Z} \nonumber
&&Z=\mbox{Tr}_{S,B}\varrho_{SB}(0)\equiv\mbox{Tr}_{S,B}\,\varrho_{eq}
=2\cosh(\beta\omega_0/2) Z_B,\\
&& Z_B=\mbox{Tr}_B e^{-\beta H_B^{(-)}}=\mbox{Tr}_B e^{-\beta
H_B^{(+)}}\eea denote the normalization functions. The explicit
expression for $Z_B$ can be derived  using Eqs.~(25) of
Ref.~\cite{MMR2012}. We will use quite a complicated form
(\ref{rho-SB-gen}) for $\varrho_{SB}(0)$ in Sec.~\ref{secIV} when
evaluating the system purity.

The correlation contribution in Eq.~(\ref{coh1}) can be rewritten
as \bea\label{cor2} \exp[\pm
i\chi(t)-\gamma_{cor}(t)]=\cos\Phi(t)\pm i {\cal
A}\sin\Phi(t).\eea The time dependent function $\Phi(t)$ is also
of the second order in the coupling parameter $g_k$; the explicit
form for $\Phi(t)$ can be found, for instance, in
Refs.~\cite{MMR2012,ourPRA}. It is clear from Eq.~(\ref{cor2})
that the correlation contribution to the coherence dynamics is
nonlinear in $|g_k|^2$, and this specific feature leads to quite
an unexpected behavior of $\langle\sigma_{\pm}(t)\rangle$ reported
in Ref.~\cite{ourPRA}

Another function, which appears in (\ref{cor2}), can be presented
as follows: \bea \label{defC} \nonumber {\cal
A}\!\!=\!\frac{\sum\limits_{m=1}^N\!\{\langle
0|\Omega_m^{\dagger}\sigma_{\pm}\Omega_m |0\rangle
e^{\beta\omega_0/2}\!-\!\langle
1|\Omega_m^{\dagger}\sigma_{\pm}\Omega_m |1\rangle
e^{-\beta\omega_0/2}\}}{\sum\limits_{m=1}^N\!\{\langle
0|\Omega_m^{\dagger}\sigma_{\pm}\Omega_m |0\rangle
e^{\beta\omega_0/2}\!+\!\langle
1|\Omega_m^{\dagger}\sigma_{\pm}\Omega_m |1\rangle
e^{-\beta\omega_0/2}\}}.\\ \eea It is seen from Eq.~(\ref{defC})
that the equality \bea\label{dual}\nonumber\sum\limits_{m=1}^N\{
\langle 0|\Omega_m^{\dagger}\sigma_{\pm}\Omega_m 0|\rangle
+\langle 1|\Omega_m^{\dagger}\sigma_{\pm}\Omega_m 1|\rangle\} \\=
\mbox{Tr}_S\left(\sigma_{\pm}\sum_{m=1}^N\Omega_m\Omega_m^{\dagger}\right)=0
\eea yields the value ${\cal A}=\coth(\beta\omega_0/2)$. The last
row in Eq.~(\ref{dual}) means that the Gram operator
(\ref{DualScheme}) has to be a diagonal one. This condition has
been shown in Ref.~\cite{ourPRA} to ensure a gradual enhancement
of the coherence at the initial time of the system evolution
(indeed, the absolute value of (\ref{cor2}) is always greater then
unity if ${\cal A}=\coth(\beta\omega_0/2)$) until the vacuum and
thermal fluctuations of the bath terminate the growth of
$\langle\sigma_{\pm}(t)\rangle$ and give rise to the decoherence
processes.

If condition (\ref{dual}) is not fulfilled, ${\cal A}$ is a
complex-valued quantity \cite{ourPRA}. In such a case, the domains
of possible coherence enhancement at some times can be
interchanged with the regions of the intensified decoherence at
other times depending on the values of the measurement state
vectors.

We are not going to study the time evolution of the system, since
the case $N>2$ can be analyzed in a similar way as it has been
done in Ref.~\cite{ourPRA} (see Eqs.~(47)-(49) of the quoted
paper), and brings nothing essentially new to our knowledge about
the the coherence dynamics. Instead, in Sec. \ref{secIV} we
concentrate on the \textit{post-measured} state of the qubit. To
this end, we will use the basic con\-di\-ti\-on (\ref{dual}) when
analyzing the system purity dependence on the measurement state
vectors $|\psi_m\rangle$, $|\varphi_m\rangle$ at $N=3$. It allows
us to reduce the number of independent Euler angles and to deal
with (at most) a three-parameter problem.

\section{Post-measurement purity of the qubit}\label{secIV}

To obtain the \textit{post-measurement} reduced density matrix
explicitly, let us calculate the trace in (\ref{rho-SB-gen}) over
the bath variables. A simple algebra yields
 \bea\label{rho-SApp}\nonumber &&
\varrho_S(0)=\mbox{Tr}_B\varrho_{SB}(0)=\sum\limits_m\omega_m|\varphi_m\rangle\langle\varphi_m|,
\\
&&\omega_m=\frac{|a_0^{(m)}|^2 e^{\beta\omega_0/2}+|a_1^{(m)}|^2
e^{-\beta\omega_0/2}}{2\cosh(\beta\omega_0/2)}. \eea Inspecting
Eq.~(\ref{rho-SApp}), one should note that the probabilities
$\omega_m$ inherited the dependence on the state of environment
via the temperature $k_B T=\beta^{-1}$, whereas the coupling
strengths $g_k$ are not involved in (\ref{rho-SApp}). Thus the
\textit{post-measurement} density matrix (\ref{rho-SApp}) depends
on the internal (temperature $T$) and external (state vectors'
amplitudes $a_i^{(m)}$) parameters. Probabilities $\omega_m$ are
normalized, $\sum_{m=1}^N\omega_m=1$, since the resolution of the
identity (\ref{Fm-NormCond}) yields $\sum_{m=1}^N
|a_0^{(m)}|^2=\sum_{m=1}^N|a_1^{(m)}|^2=1$.

One of the measures of the ``mixedness'' (or the lack of
information about a system) is the so-called purity ${\cal P}$ of
the system's state \cite{ourPRA,Luczka,2inDajka},
\bea\label{purity} {\cal P}=\mbox{Tr}_S[\varrho_S(0)]^2. \eea
Obviously $1/2\le {\cal P} \le 1$  with ${\cal P} = 1$ for a pure
state. Taking into account that the state vectors
$|\varphi_m\rangle$ do not form the orthogonal basis, and denoting
\bea\label{C_ij} C_{mn}=\langle\varphi_m|\varphi_n\rangle, \eea
one can write down the expression for the
\textit{post-measurement} purity of the qubit as follows:
\bea\label{purity1} {\cal P}=\sum\limits_{m,n=1}^N
|C_{mn}|^2\omega_m\omega_n. \eea Using expressions
(\ref{a-qubit})-(\ref{b-qubit}) for the measurement state vectors,
one can obtain the values $|C_{mn}|^2$ in the explicit form,
\bea\label{C_ijGen}\nonumber
|C_{mn}|^2=\cos^2\left(\frac{\phi_b^{(m)}-\phi_b^{(n)}}{2}\right)
\cos^2\left(\frac{\theta_b^{(m)}-\theta_b^{(n)}}{2}\right)
\\+\sin^2\left(\frac{\phi_b^{(m)}-\phi_b^{(n)}}{2}\right)
\cos^2\left(\frac{\theta_b^{(m)}+\theta_b^{(n)}}{2}\right). \eea

Our main goal in this Section is to investigate the extremal
values of $\omega_m$, yielding the minima of the bilinear form
(\ref{purity1}), and to analyze stabilities of these minima with
respect to variation of the internal and external parameters. A
motivation of our researches is the following. Since $\sigma_3$ is
the integral of motion, see Eq.~(\ref{H}), the measurement
settings should avoid the situations when the minima ${\cal
P}_{min}$ of the qubit purity are realized. Moreover, it will be
shown later that at some conditions ${\cal P}_{min}=1/2$ that
means a completely decohered system with equal population of the
levels. If the initial value of the coherence turns out to be
zero, $\langle\sigma_{\pm}\rangle=0$, no dynamic purification
\cite{Luczka,ourPRA} of the system is possible. It is clear, that
such a situation is extremely undesirable in the quantum control
problems \cite{measur,q-control}, and the ways to exclude it
should be found.

Let us look at the extremum of the system purity from the
mathematical point of view. In fact, we deal with a conditional
minimum problem of the bilinear form (\ref{purity1}), generated by
a symmetric matrix $||\textbf{C}||\equiv|C_{mn}|^2$, with an
additional constrain $\sum_{m=1}^N\omega_m=1$. It is also obvious,
that only located within the interval $[0,1]$ extremal values of
the probabilities $\omega_n^{(ext)}$ have a physical mean\-ing.
Having solved the conditional extremum problem for
(\ref{purity1}), after a simple algebra one obtains
\bea\label{Wextr}
0\le\omega_n^{(ext)}=\frac{\sum\limits_{m=1}^N||\textbf{C}^{-1}||_{mn}}{\sum\limits_{m,n=1}^N
||\textbf{C}^{-1}||_{mn}}\le 1, \eea where $||\textbf{C}^{-1}||$
denotes the matrix inverse to $||\textbf{C}||$. Substituting
(\ref{Wextr}) in Eq.~(\ref{purity1}), we obtain the minimal value
of the initial purity of the qubit, \bea\label{Pmin} {\cal
P}_{min}= \frac{1}{\sum\limits_{m,n=1}^N
||\textbf{C}^{-1}||_{mn}}.\eea

It is seen from (\ref{Wextr})-(\ref{Pmin}) that locations of
$\omega_n^{(ext)}$ as well as the value of ${\cal P}_{min}$ are
completely defined by the absolute values of the scalar products
of the measurement state vectors $|\varphi_m\rangle$ in the
two-dimensional Hilbert space. This general result for any number
of the outcomes has, however,  two distinguishing features: while
at $N=2$ the extremal values of probabilities are always equal to
$\omega_1^{(ext)}=\omega_2^{(ext)}=1/2$, the character of
locations of $\omega_m^{(ext)}$ at $N=3$ is much more diverse.

First of all, if any of $\omega_m^{(ext)}$ goes beyond the
interval $[0,1]$, the purity $(\ref{purity1})$ does not have a
global minimum, and, consequently, the system cannot be completely
decohered.

Secondly, if the measurement state vectors $|\varphi_m\rangle$ are
chosen in such a way that all $|C_{mn}|$ are the same at $m\ne n$,
the extremal values of probabilities $\omega_m^{(ext)}$ can be
shown to be equal to 1/3 for any $m$. Such a measurement setting
resembles the case $N=2$ in the sense that probabilities
$\omega_m^{(ext)}$ cease to depend on temperature, see
Eq.~(\ref{rho-SApp}). We will discuss this important fact as well
as the measurement settings, leading to the above mentioned
location of $\omega_m^{(ext)}$, later in more details.

At last but not least, the extremum conditions
(\ref{Wextr})-(\ref{Pmin}) require the parameters of the
measurement state vectors $|\psi_m\rangle$ to be chosen
self-consistently. In particular, it follows from
Eqs.~(\ref{a-qubit})-(\ref{b-qubit}) and (\ref{rho-SApp}) that
Euler angles $\theta_a^{(m)}$ have to obey the equation
\bea\label{cond-thetA} \cos\theta_a^{(m)}=\left( 1-N
\omega_m^{(ext)} \right)\coth
\left(\frac{\beta\omega_0}{2}\right),\quad N=3.\eea

Having excluded the above mentioned case $\theta_a^{(m)}=\pi/2$
(when $\omega_m^{(ext)}=1/3$) from consideration, one can ask the
following question: what a researcher performing the nonselective
measurements has to do in order to purify the system, which could
fall down into the most mixed state due to an inappropriate
measurement setting? An answer turns out to be quite surpris\-ing:
for instance, to heat it. Indeed, though the coupling strength is
not involved explicitly in the reduced density matrix
(\ref{rho-SApp}), there is a spin-phonon interaction in the
composite ($S+B$) system after the measurement, which defines the
\textit{post-measurement} density matrix (\ref{rho-SB-gen}). Thus
one can always re-prepare (via interaction with the environment)
the qubit in the state given by Eq.~(\ref{rho-SApp}) but with a
different temperature: formally, one just takes the trace over the
bath variables in the generic \textit{post-measurement} density
matrix (\ref{rho-SB-gen}) with the altered temperature. In this
context, the temperature variation can be regarded as a
``repeated'' measurement (or, more precisely, as an additional
action) which is performed over the thermal bath rather than the
spin.

Thus by heating the system right after the measurement, one can
force the right-hand-side of Eq.~(\ref{cond-thetA}) to go beyond
the limit $|\cos\theta_a^{(m)}|\le 1$. Once this happens, no
selection of $\theta_a^{(m)}$ is possible at fixed parameters of
the measurement state vectors $|\varphi_m\rangle$ to provide the
system state with a global minimum of its purity. \footnote{It is
interesting \cite{ourPRA}, that the heating reduces both
$\langle\sigma_{\pm}\rangle$ and the negative value of
$\gamma_{cor}(t)$. Thought the first factor reduces the system
purity, the second one increases it; the joint effect of both
processes is a short-lasting dynamic purification of the qubit.} .
One can say that the temperature variation can serve as a source
of ``pulling'' of the system up from the most mixed state with
subsequent purification. This case completely differs from the
Neumann-L\"uders projection at $N=2$, when the global minimum of
${\cal P}$ is realized iff $\cos\theta_a^{(m)}=\pi/2$ (and
$\omega_m^{(ext)}=1/2$). The values of the initial coherence
$\langle\sigma_{\pm}\rangle$ and mean population of the levels
$\langle\sigma_{3}\rangle$ turn out to be temperature-independent
at $\cos\theta_a^{(m)}=\pi/2$ (see Eqs.~(61) and (67) of
Ref.~\cite{ourPRA}), and a complete loss of the system purity
cannot be avoided by the temperature change.

Now let us perform a brief review of some measurement settings,
which lead to the most mixed \textit{post-measurement} state. In
doing so, we use the results of Appendix A, where different
measurement scenarios ensuring condition (\ref{dual}) of the
gradual coherence enhancement at the initial stage of the system
evolution have been considered. As we have already mentioned, this
allows to reduce our study to (at most) the three-parameter
problem.

In the \underline{\textbf{case 1}}, the squared modulus of the
scalar pro\-ducts of the measurement state vectors
$|\varphi_m\rangle$ is expressed by Eq.~(\ref{Cijcase1}). Having
inserted these values in Eq.~(\ref{Pmin}), we obtain that ${\cal
P}_{min}=1/2$ for any $\theta_b^{(1)}$, $\theta_b^{(2)}$ (note
also that the third polar angle $\theta_b^{(3)}$ cannot be chosen
in an arbitrary way and has to satisfy Eqs.~(\ref{mn})). At the
measurement settings of case 1 we obtain the most mixed system: a
completely decohered and with equal populations of the levels.
However, if $\theta_a^{(m)}\ne\pi/2$, this global minimum of the
system purity is dependent on temperature, see
Eq.~(\ref{cond-thetA}). Hence, the system can be pulled up from
this minimal value of ${\cal P}$ by changing temperature right
after the measurement. In Sec. \ref{secV} we consider another way
of the system purification by series of the repeated measurements,
when the second measurement occurs immediately after the first
one.

At $\theta_a^{(m)}=\pi/2$, the extremal values of the probability
$\omega_m^{(ext)}$ are equal to $1/3$ and do not depend on
temperature. Such a situation occurs at the conditions described
in Appendix B. Thus one has to avoid a simultaneous realization of
such measurement settings since the system is completely decohered
and cannot be purified by nothing except another measurement.

In the \underline{\textbf{case 2}}, the squared modules of the
scalar products of measurement state vectors $|\varphi_m\rangle$
are expressed by Eqs.~(\ref{C_ij_case2n0})-(\ref{C_ij_case2n1}).
Having inserted these values in Eq.~(\ref{Pmin}), we obtain that
${\cal P}_{min}=1/2$ at any $\theta_b^{(1)}$ . Again, this global
minimum of the system purity is temperature-dependent if
$\theta_a^{(m)}\ne\pi/2$. However, it ceases to depend on the
system temperature if $\theta_b^{(1)}$ obeys (\ref{CequalCase2}),
and we have a close analogy with the case 1 except that now we are
dealing with the one-parameter problem.

In the \underline{\textbf{case 3}}, no measurement settings can
give the condition $\omega_m^{(ext)}=1/3$ as it is claimed at end
of Appendix B, and the global minimum of the system purity is
always ${\cal P}_{min}> 1/2$. Hence, there is no danger to occur
at the totally decohered state with an equal population of the
levels after an ``inappropriate'' choice of the experimental
setting. In the case 3, all the extremal values of the
probabilities (\ref{rho-SApp}) depend on temperature, and the
system can always be purified by chang\-ing temperature. The
results of case 3 resemble those (see Eqs.~(66)-(67) of
Ref.~\cite{ourPRA}) of the measurement settings during the von
Neumann-L\"uders projection. The only difference is that the
global minimum ${\cal P}_{min}>1/2$ in Ref.~\cite{ourPRA} is
realized at $\omega_m^{(ext)}=1/2$ and is temperature-independent.

It is also to be stressed that in the case 3 we deal with a
three-parameter problem (see the end of Appendix A). However, the
parameters $\Delta_{12}$, $\Delta_{13}$ and $\theta_b^{(1)}$
cannot be chosen in an arbitrary way: according to
Eqs.~(\ref{Acond_case3}), there is a ``window'' of the allowed
values of Euler angles ensuring the condition (\ref{dual}).

Now some quantitative estimations to what extent the system could
be purified by the temperature variation are in order. Before
doing so, let us note that in the case 1 the rate of the
temperature change could be very small since no dynamics of the
system coherence is possible. Indeed, when condition (\ref{dual})
is satisfied, the initial coherence is determined by the equation
\bea\label{coh1(0)}\langle\sigma_{\pm}\rangle=\sum\limits_{m=1}^N
\langle
0|\Omega^{\dagger}_m\sigma_{\pm}\Omega_m|0\rangle\tanh(\beta\omega_0/2)
\eea and cannot be affected by the temperature variation if,
initially, $\sum_{m=1}^N \langle
0|\Omega^{\dagger}_m\sigma_{\pm}\Omega_m|0\rangle=0$. In such a
case, the system purification occurs only via the re-distribution
of the levels population, which is time-independent, because
$\sigma_3$ is the integral of motion, see Eq.~(\ref{H}).

Thus we consider the case 3 of the measurement settings because it
is more interesting from the practical point of view. Suppose that
we chose the subset of the measurement state vectors
$|\varphi_m\rangle$ and composed the accompanying subset
$|\psi_m\rangle$ using the self-consistency condition
(\ref{cond-thetA}) at the fixed bath temperature $T^{\star}$. At
this temperature, the system is definitely in the most mixed
state. After that, we alter the environment temperature $T$ only
and measure the system purity. The reaction of the system purity
to such temperature variations is depicted in Fig.~1.
 \begin{figure}[htb]
\centerline{\includegraphics[height=0.3\textheight,angle=0]{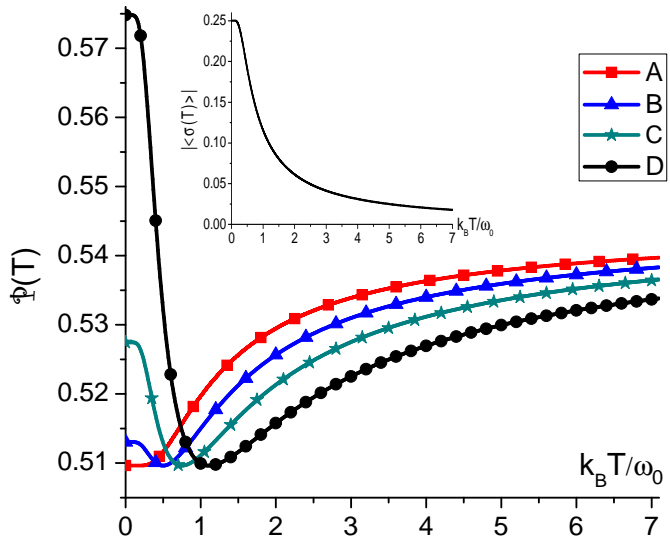}}
\caption{Change of the system purity ${\cal P}(T)$ with the bath
temperature $T$. The values of the Euler angles $\theta_a^{(m)}$
are defined at the fixed temperatures $k_B
 T^{\star}/\omega_0=0$ (A), 0.5 (B), 0.75 (C) and 1.11 (D) due to the self-consistency condition (\ref{cond-thetA})
 and are not changed after the measurement.
The other Euler angles $\theta_b^{(1)}=4\pi/5$,
$\Delta_{12}=7\pi/5$, $\Delta_{13}=3\pi/4$ are taken from the
allowed ``window'' of values (\ref{Acond_case3}). In the inset:
the temperature dependence of the initial coherence
$|\langle\sigma(T)\rangle|\equiv|\langle\sigma_{\pm}(T)\rangle|$
at the parameters of the choice D.}
\end{figure}
It is seen that the purity minima are reached at temperatures
$T^{\star}$ as it should be due to the self-consistency condition
(\ref{cond-thetA}). However, both reduction and increase of the
thermal bath temperature $T$ ``pull'' the system up from the most
mixed state. A detailed analysis of the probabilities
(\ref{rho-SApp}) shows that in the zero-temperature limit the
system purity is defined by the ground state amplitudes according
to \bea\label{purity_low} {\cal P}(T=0)=\sum\limits_{m=1}^N
|C_{mn}|^2|a_0^{(m)}|^2|a_0^{(n)}|^2, \eea whereas in the
high-temperature limit the purity tends to the value
\bea\label{purity_high} {\cal P}(\beta\omega_0\ll
1)=\frac{1}{9}\,\sum\limits_{m=1}^N |C_{mn}|^2.\eea It is
interesting that in this limit all the probabilities
(\ref{rho-SApp}) tend to $\omega_m=1/3$, which coincide with the
extremal values $\omega_m^{(extr)}$ obtained in the case 1 at
$\theta_a^{(m)}=\pi/2$. It can be concluded from Fig.~1 that at
low values of $T^{\star}$ heat\-ing is favorable for the system
purification while at high values of $T^{\star}$ cooling yields
less mixed states. The highest considered value of $k_B
T^{\star}/\omega_0=1.11$ corresponds to the maximal temperature,
at which the self-consistency condition (\ref{cond-thetA}) still
holds true at the chosen state vectors $|\varphi_m\rangle$, and
the system is allowed to be at the global minimum of its purity.
The general conclusion, which could be done after inspection of
Fig.~1, consists in a remarkable fact that the system purity can
be increased even by several tens of percents by temperature
variation.

In the inset of Fig.~1, we depicted the temperature dependence of
the initial system coherence
$|\langle\sigma(T)\rangle|\equiv|\langle\sigma_{\pm}(T)\rangle|$
obtained at the parameters of the choice D. It is seen that the
initial coherence follows the $\tanh(\beta\omega_0/2)$ rule and is
never equal to zero.

Since the system-environment correlations play a constructive role
at the early stage of the system evolution, and the coherence
enhances at small times \cite{ourPRA}, the initial state
re-preparation by the temperature variation can be considered as
an additional way to obtain a purer system. It should be stressed
that it is possible to obtain a behavior of ${\cal P}(T)$ similar
to that one presented in Fig.~1 even in the cases 1 and 2, but the
minima of purity would be equal to 1/2. As we have already
mentioned, no dy\-namic purification of the system could be
possible in such a case.

We conclude this Section by consideration of some limiting cases,
when the number of vectors in the subset $|\varphi_m\rangle$ is
less than 3. At the same time, we allow the dimension of the
vector subset $|\psi_m\rangle$ to be equal to $N=3$.

If $|\varphi_m\rangle\equiv|\varphi\rangle$ for any $m=\{1,2,3\}$,
we obtain the pure state of the qubit after the nonselective
measurement (see Eq.~(\ref{rho-SApp})) since the probabilities
$\omega_m$ are normalized to the unity. Here we are in a close
resemblance with the measurement settings (62) of
Ref.~\cite{ourPRA}, yielding the value (63) of the correlation
contribution $\gamma_{cor}(t)$ to the generalized decoherence
function $\tilde\gamma(t)$. In the Ohmic case \cite{MMR2012}, the
system coherence starts to decrease monotonically from the value
(65) to zero at $t\to\infty$.

At $|\varphi_1\rangle\ne|\varphi_2\rangle=|\varphi_3\rangle$, the
situation is much more interesting. On the one hand, we have a
mapping to the von Neumann-L\"uders projection scheme
\cite{ourPRA} since \bea\label{map-to-N2}
\varrho_S(0)=\omega_1|\varphi_1\rangle\langle\varphi_1|+
\bar{\omega}_2|\varphi_2\rangle\langle\varphi_2|,\qquad
\bar{\omega}_2=\omega_2+\omega_3. \eea
To satisfy the condition
(\ref{dual}) of the gradual coherence enhancement at small times,
the measurement state vectors $|\varphi_1\rangle$ and
$|\varphi_2\rangle$ have to be chosen to be orthonormal (see
Eq.~(55) of Ref.~\cite{ourPRA}) or to obey Eqs.~(66). However, no
selections of the above mentioned state vectors can ensure the
global minimum of the system purity, since the requirement
$\omega_1^{(ext)}=\omega_2^{(ext)}=1/2$ (related to equalities
$|a_0^{(1)}|^2=|a_1^{(1)}|^2=1/2$) contradicts the normalization
conditions (\ref{psi-psi}). Hence, at the case considered the
system can be only in the state with a local minimum of purity,
and we should not be afraid to obtain the completely decohered
system after the measurement.

\section{The system response to the repeated nonselective
measurements \label{secV}}

In the previous section we discussed the influence of temperature
variation on possible purification of the open quantum system. The
temperature was supposed to have been changed right after the
nonselective measurement.

Let us now consider a situation, when the system is subjected to
the second nonselective measurement, which occurs immediately
after the first one. We can also imagine a more general situation,
when the system is subjected to the series of the repeated
measurements at the initial instant of time $t=0$.

If such a measurement scheme is applied, the initial density
matrix $\varrho'_{SB}(0)$ of the composite system right after the
first measurement obeys the equation \bea\label{eq1prime}
\varrho'_{SB}(0)=\sum\limits_{m=1}^N
\Omega'_m\varrho_{SB}(0)\Omega'^{\dagger}_m.\eea This equation is
analogous to the basic relation (\ref{rho-SB-0}), where the
equilibrium density matrix $\varrho_{eq}$ is replaced by the
post-measurement one $\varrho_{SB}(0)$. The ``primed'' operators
$\Omega'_m$ are defined in the way similar to (\ref{Omega-m-gen}),
where the state measurement vectors are replaced accordingly
$|\varphi_m\rangle\rightarrow|\varphi'_m\rangle$ and
$|\psi_m\rangle\rightarrow|\psi'_m\rangle$.

Thus, taking the trace in (\ref{eq1prime}) over the bath
variables, one can obtain the expression for the system density
matrix $\varrho'_S(0)$ after the second measurement,
 \bea\label{rho-S-prime}
\varrho'_S(0)=\sum\limits_{n=1}^N\tilde\omega_n|\varphi'_n\rangle\langle\varphi'_n|,
\eea where the probabilities $\tilde\omega_n$ after the second
measurement are expressed \bea\label{w-tilde}
\tilde\omega_n=\sum\limits_{m=1}^N |D_{nm}|^2\omega_m,\qquad
D_{nm}=\langle\psi'_n|\varphi_m\rangle \eea via the scalar
products of the ``old'' $|\varphi_m\rangle$ and ``new'' (primed)
$|\psi'_m\rangle$ measurement state vectors.

The system purity right after the second measurement can be
presented in the form similar to (\ref{purity1})
\bea\label{purity-prime} {\cal
P}'=\sum\limits_{m,n=1}^N|C'_{mn}|^2\tilde\omega_m\tilde\omega_n,
\eea where the matrix elements $C'_{mn}$ are nothing but those of
(\ref{C_ij}) defined on the ``new'' subset $|\varphi'_m\rangle$.

It would be constructive to consider some different choices of the
new measurement state vectors $|\psi'_m\rangle$ and
$|\varphi'_m\rangle$ in relation to the change of the system
purity.

\subsection{Square root measurements}

Suppose that the open quantum system after the first measurement
was characterized by the density matrix (\ref{rho-SB-gen}). Now
let us construct the ``new'' measurement state vectors
$|\psi'\rangle$ according the general rule \bea\label{psi-SRM}
|\psi'_n\rangle=\sqrt{\omega_n}\varrho_S^{-1/2}(0)|\varphi_n\rangle.
\eea Here, $\varrho_S^{-1/2}(0)$ denotes the generalized inverse
of the operator $\varrho_S^{1/2}(0)$. The operator $\varrho_S(0)$
as well as its square root are invertible since the rank $n$ of
the density matrix $\varrho_S(0)$ is equal to the dimension of the
Hilbert space ${\cal H}$ of the measurement state vectors.
Moreover, its is a straightforward matter to calculate
$\varrho^{-1/2}_S(0)$ explicitly for $2\times 2$ matrix of the
initial density operator of the system.

The measurement state vectors $|\psi'_n\rangle$, introduced by
Eq.~(\ref{psi-SRM}), are known to minimize \cite{SRM,SRM1} the
squared error $R$, determined by \bea\label{R}
R=\sum\limits_{n=1}^N
\omega_n||\,|\varphi_n\rangle-|\psi'_n\rangle||^2. \eea
The
definition (\ref{R}) of the square error $R$ has to be considered
in terms of the Hilbert-Schmidt norm of some specific operators
constructed on the measurement state vectors $|\psi'_n\rangle$ and
$|\varphi_n\rangle$ (see Ref.~\cite{SRM} and the references herein
for more details). Such schemes with the effects
$F'_n=|\psi'_n\rangle\langle\psi'_n|$, which realize the
least-square root $R$, are known as \textit{the square root
measurements (SRM)} and are the subject of thorough investigation
in the quantum measurement theory \cite{SRM,SRM1,Holevo1979}.

Now let us look at the probabilities (\ref{w-tilde}), when the
basic subset of the measurement state vectors $|\psi'_m\rangle$ is
taken in the SRM form (\ref{psi-SRM}). It is obvious that
\bea\label{w-identity} && \tilde\omega_n=\sum\limits_{m=1}^N
|\langle\varphi_n|\varrho_S^{-1/2}(0)|\varphi_m\rangle|^2\omega_m\omega_n\\
\nonumber
&&=\omega_n\large\langle\varphi_n|\varrho_S^{-1/2}(0)\underbrace{\!\sum\limits_{m=1}^N\omega_m|\varphi_m
\large\rangle\large\langle\varphi_m|}_{\rho_S(0)}\varrho_S^{-1/2}(0)|\varphi_n\large\rangle\equiv\omega_n.
\eea Thus, if one chooses $|\psi'_n\rangle$ in the SRM form, then
the system purity (\ref{purity-prime}) after the second
measurement is expressed via the probabilities $\omega_n$,
obtained after the first measurement. Suppose that the system
after the first measurement is prepared in the most mixed state;
this minimal va\-lue of the purity is realized at
$\omega_n^{(ext)}$ (see Eq.(\ref{Wextr})), and the
self-consistency condition (\ref{cond-thetA}) is obeyed. Now let
us consider two possible selections of the subset
$|\varphi'_m\rangle$ of measurement state vectors:
\begin{enumerate}
\item The subsets of the the ``new'' and ``old'' vectors are
different, $|\varphi'_m\rangle\ne|\varphi_m\rangle $.  After the
second measurement, the extremum points $\tilde\omega^{(ext)}_m$
are shifted from their previous values, whereas the probabilities
$\tilde{\omega}_m=\omega_m$ remain unchanged. Thus, the
self-consistency condi\-tion (\ref{cond-thetA}) is now violated,
and the system purity is no longer minimal; \item The subsets of
the ``new'' and ``old'' vectors are identical,
$|\varphi'_m\rangle\equiv|\varphi_m\rangle $. At this condition,
``old'' and ``new'' purities are the same since the state of the
system does not change after the second measurement.
\end{enumerate}
Hence, the result of the choice (\ref{psi-SRM}) of the ``new''
state vectors $|\psi'_m\rangle$ can be ambiguous: in the case 1 we
can assert that the system is not in the global minimum of purity
after the second measurement, while in the case 2 the system
remains in the same most mixed state, and the repeated measurement
is useless from the viewpoint of the system purification.
\subsection{Ordinary measurements}

Now suppose that the definition (\ref{psi-SRM}) is not applied,
and the measurement state vectors $|\psi'_m\rangle$ are chosen to
be arbitrary ones. Once again, let us suppose that the system was
at the point of the global minimum of the purity after the first
measurement and consider various choices of the subsets
$|\varphi'_m\rangle$:
\begin{enumerate}
\item The ``new'' and ``old'' subsets are identical,
$|\varphi'_m\rangle=|\varphi_m\rangle$. In such a case, the system
purity is defined by Eq.~(\ref{purity-prime}), where the matrix
elements are not changed,  $C_{mn}=C'_{mn}$. Contrary to the case
1 from the previous subsection, the extremum points
$\omega_m^{(ext)}$ remain unaltered after the second measurement.
However, the self-consistency condition (\ref{cond-thetA}) is
violated since the system purity is determined by the ``new''
probabilities $\tilde\omega_m$ defined on the modified subset
$|\psi'_m\rangle\ne|\psi_m\rangle$. Thus, the system is no longer
in the global minimum of purity after the second measurement;
\item Let us, in addition to the equality
$|\varphi'_m\rangle=|\varphi_m\rangle$, apply another condition,
$|\psi'_m\rangle=|\varphi_m\rangle$. The matrix elements $D_{mn}$,
which define the ``new'' probabilities $\tilde\omega_m$ (see
Eq.~(\ref{w-tilde})), are now expressed as
$D_{mn}=\langle\varphi_m|\varphi_n\rangle$. Suppose, we are
perform\-ing the Neumann-L\"uders projective measurements with
$N=2$ and choose the vectors $|\varphi_1\rangle$ and
$|\varphi_2\rangle$ to be orthogonal to each other (such a
selection is one of the way to ensure the gradual growth of the
system coherence at the initial stage of its evolution
\cite{ourPRA}). At such a choice, the matrix elements transform
into the Kronecker symbols, $D_{mn}=\delta_{mn}$. As a result, the
``new'' probabilities remain the same as the ``old'' ones,
$\tilde\omega_m=\omega_m$. It means that nothing has been changed
in the state of the system, and its purity is the same, ${\cal
P'}={\cal P}$.

\hspace{3mm} In all other cases (including the measurements on the
\textit{ overcomplete} basis), $D_{mn}\ne\delta_{mn}$, thus
$\tilde{\omega}_m$ $\ne$ $\omega_m$, and we actually come back to
the item 1; \item

Now let us perform the second repeated measurement at the
completely modified subsets $|\varphi'_n\rangle\ne
|\varphi_n\rangle$ and $|\psi'_n\rangle\ne |\psi_n\rangle$. It is
obvious that the extremum probabilities before and after the
second measurement are different,
$\tilde\omega^{(ext)}_n\ne\omega_n^{(ext)}$. To ensure that the
system would not drop again into the global minimum of purity, it
is necessary to violate the self-consistency condition
(\ref{cond-thetA}). Using definitions (\ref{Wextr}) and
(\ref{w-tilde})-(\ref{purity-prime}), one can express this
requirement as an inequality \bea\label{violateWext}
\frac{\sum\limits_{m,p=1}^N ||\textbf{D}^{-1}||_{nm}||\textbf{
C}'^{-1}||_{mp}}{\sum\limits_{m,p=1}^N||\textbf{C}'^{-1}||_{mp}}\ne
\frac{\sum\limits_{p=1}^N ||\textbf{
C}^{-1}||_{nm}}{\sum\limits_{p,m=1}^N||\textbf{C}^{-1}||_{nm}},
\eea where $||\textbf{D}^{-1}||$ denotes the matrix inverse to
$||\textbf{D}||\equiv |D_{mn}|^2$. The matrix $||\textbf{C}'||$ is
defined by (\ref{C_ij}) on the ``new'' (primed)
subset of the
measurement state vectors $|\varphi'_m\rangle$.

\hspace{3mm} If condition (\ref{violateWext}) is satisfied, the
system is not in the global minimum of its purity for sure.
However, to clarify how pure the system is after the second
measurement in comparison with its state after the first one, an
additional study is necessary since the global minimum ${\cal
P}_{min}$ can be higher than the non-extremal value ${\cal P}'$.

\end{enumerate}

To conclude this section, let us make a brief summary of the above
considered cases from the viewpoint of a possible system
purification:
\begin{itemize}
\item The measurement schemes discussed in the
\underline{\textbf{case 1, subsection A}}, as well as in the
\underline{\textbf{case 1, subsection B}}, are unambiguously
pro\-mis\-ing since the system after the second measurement is
pulled up from the global minimum of its purity. The same is true
about the measurements on the \textit{overcomplete} basis
described in the \underline{\textbf{case 2, subsection B}}; \item
The scheme considered in the \underline{\textbf{case 2, subsection
A}}, is useless from this viewpoint since the system purity is not
changed after the repeated measurement. The same is true about the
special kind of the Neumann-L\"uders projection scheme discussed
in the \underline{\textbf{case 2, subsection B}}; \item The
\underline{\textbf{case 3, subsection B}}, cannot give us
unambiguous answer about the system purification and has to be
studied in more details. However, at this measurement scheme the
system is definitely not in the global minimum of its purity as
compared with the first measurement.
\end{itemize}

\section{Conclusions\label{secVI}}

In this paper, we study the initial state preparation of the qubit
after the nonselective measurement, when the number of outcomes
$N$ is greater than two. In such a scheme, the basis of the
measurement state vectors is known to be overcomplete
\cite{Holevo2001,SRM,Holevo1979}, and the vectors are not
orthonormal. To obtain the explicit results of practical
importance, we have chosen the exactly solvable dephasing model
\cite{Luczka,MMR2012} and applied the conditions for the
post-measurement state to ensure a gradual coherence enhancement
\cite{ourPRA} at the initial stage of the system evolution.

We show that at some special choices of parameters the system
after the nonselective measurement can be prepared in the most
mixed state. This case is strongly undesirable from the viewpoint
of the quantum control \cite{measur} since the system could remain
in the completely decohered state, and no dynamic purification
\cite{ourPRA} of the qubit is possible. It is pointed out that, in
general, the minimal value ${\cal P}_{min}$ of the system purity
depends on the temperature of the thermal bath. Though at some
measurement settings ${\cal P}_{min}$ can be equal to 1/2, and the
system is totally decohered, there is still a possibility to
purify the state of the qubit by changing temperature and,
consequently, by making a re-distribution of populations of the
levels.

We have also considered a more interesting case (from the
practical viewpoint) when ${\cal P}_{min}>1/2$. At such
measurement settings, both dynamical and temperature-controlled
purifications are possible. We show that by alter\-ing the
environment temperature after the nonselective measurement, it is
possible to increase the system purity even by tens of percents.
If the composite ($S+B$) system was initially prepared at low
temperature, and the qubit was in the most mixed state, then
heating is preferable. Other\-wise, cooling yields a purer system.

However, there are also some other measurement settings, when the
most mixed state of the system cannot be influenced by
temperature, and the system purification is possible only after
another measurement. In this case, a character of the minimum of
${\cal P}$ is very similar to that in the von Neumann-L\"uders
projection scheme, when the most mixed state of the qubit is
always independent of temperature. We have listed the parameters
options leading to such a scenario and noted that the measurement
settings of this kind have to be avoided from the viewpoint of the
problems of quantum control.

We have also considered two measurements, when the numbers of the
state vectors in two basic subsets are different, and found out
some similar and different features as compared to the case
studied earlier \cite{ourPRA}.

One of the ways to maintain the steady coherence of the system by
a sequence of the repeated selective measurements due to the
quantum Zeno effect is reported recently in
Ref.~\cite{Chaudhry14}. In our paper, we proceed in a bit
different way and consider the sequence of the non\-selective
measurements performed instantly at the initial time $t=0$ of the
system preparation. It turns out that some kinds of the
measurement settings during such a repeated quantum measurements
can be perspective from the viewpoint of an eventual system
purification at $t=0$, while the other schemes do not allow to
pull the system away from the point of global maximum of its
mixedness. The action of some other schemes considered is
ambiguous from the viewpoint of qubit purification, and it
requires an additional study.

It would be interesting to verify our measurement schemes on more
realistic systems, which allow not only the decoherence
\cite{27inPRA12,28inPRA12} but also the population decay
\cite{PRA2010,Ban,severalW}. However, it is a challenging problem
because in this case the model can be no longer exactly solvable.

\appendix

\section{Measurement settings that ensure
condition (\ref{dual})}
\renewcommand{\theequation}{A.\arabic{equation}}
\setcounter{equation}{0}

To provide an enhancement of the system coherence, the Gram
operator (\ref{DualScheme}) in accordance with Eq.~(\ref{dual})
has to be diagonal. Thus the following condition
 \bea\label{ReIm0}
\sum\limits_{m=1}^3\exp(i\phi_b^{(m)}) \sin\theta_b^{(m)}=0 \eea
has to be ensured.
 Equating both the real and the imaginary parts of
(\ref{ReIm0}) to zero and expressing the third polar angle trough
the remaining ones, it is possible to rewrite (\ref{ReIm0}) as
follows: \bea\label{non-theta3} \nonumber &&
\sin\left(\phi_b^{(1)}-\phi_b^{(3)}\right)\sin\theta_b^{(1)}-\sin\left(\phi_b^{(2)}-\phi_b^{(3)}\right)\sin\theta_b^{(2)}=0,
\\
&& \cos\phi_b^{(3)}\ne 0. \eea Now we consider three different
options when condition (\ref{non-theta3}) is satisfied.
\vspace{-2mm}
\subsection{The case $\sin\theta_b^{(1)}\ne 0$, $\sin\theta_b^{(2)}\ne
0$, $\phi_b^{(m)}-\phi_b^{(n)}=\pi k$. \label{case1}}
\vspace{-2mm}

These conditions assume that $\phi_b^{(1)}-\phi_b^{(3)}=\pi k$,
$\phi_b^{(2)}-\phi_b^{(3)}=\pi l$, and
$\phi_b^{(1)}-\phi_b^{(2)}=\pi (k-l)$, $\{k,l\}=\{0,1\}$. Having
expressed two azimuthal angles via the third one and taking into
account that \bea\label{0pi} 0\le\theta_b^{(m)}\le\pi,\eea one
gets the following equations for $\theta_b^{(3)}$ (at the
additional assumption $\theta_b^{(1)}>\theta_b^{(2)}$):
\bea\label{mn}
&&\nonumber\sin\theta_b^{(3)}=\sin\theta_b^{(1)}-\sin\theta_b^{(2)}\quad
\mbox{at}\quad k=1,\, l=0,\\
&&\sin\theta_b^{(3)}=\sin\theta_b^{(1)}+\sin\theta_b^{(2)}\quad
\mbox{at}\quad k=l=1.\eea Eqs.~(\ref{mn}) impose restrictions on
the possible values of the polar angles
$\theta_b^{(1)},\theta_b^{(2)}$ due to inequality (\ref{0pi}).

The values (\ref{C_ijGen}) in the case 1 can be expressed as
\bea\label{Cijcase1}
|C_{mn}|^2=\cos^2\frac{\left(\theta_b^{(m)}\mp\theta_b^{(n)}\right)}{2}.
\eea The upper sign in Eq.~(\ref{Cijcase1}) has to be taken for
$\{m=1, n=2\}$ at $l=1$ and for $\{m=2, n=3\}$ at
 $l=0$; the lower sign has to be chosen for $\{m=1, n=3\}$ as
well as for pairs
 $\{m=1, n=2\}$ at $l=0$ and $\{m=2, n=3\}$ at
 $l=1$.

In fact, in the case 1 we deal with a two-parameter problem at the
nonselective measurement, where only two polar angles
$\theta_b^{(1)}$ and $\theta_b^{(2)}$ are necessary. There is also
a ``window'' of the allowed values for the third polar angle
$\theta_b^{(3)}$ defined by Eqs.~(\ref{0pi})-(\ref{mn}).
\vspace{-3mm}
\subsection{The case $\sin\theta_b^{(1)}\ne 0$,
$\sin\theta_b^{(2)}=0$.\label{case2}} \vspace{-3mm} This case
implies that $ \theta_b^{(2)}=\pi k$, $k=\{0,1\}$, and
$\phi_b^{(1)}-\phi_b^{(3)}=\pi$. Taking into account
Eqs.~(\ref{C_ijGen}), (\ref{ReIm0}) and having ex\-pressed
$\theta_b^{(3)}$ via $\theta_b^{(1)}$, one can obtain after some
algebra: \bea\label{C_ij_case2n0}
&&|C_{12}|^2=\cos^2(\theta_b^{(1)}/2),\\[1ex]
\nonumber
&&|C_{13}|^2=\Biggl[\begin{array}{ccc}\cos^2\theta_b^{(1)}&
\mbox{at}&\theta_b^{(3)}=\theta_b^{(1)},\\0 &
\mbox{at}&\qquad\theta_b^{(3)}=\pi-\theta_b^{(1)},\end{array}\\
\nonumber
&&|C_{23}|^2=\Biggl[\begin{array}{ccc}\cos^2(\theta_b^{(1)}/2)&
\mbox{at}&\!\!\theta_b^{(3)}=\theta_b^{(1)},\\
\sin^2(\theta_b^{(1)}/2)&
\mbox{at}&\quad\theta_b^{(3)}=\pi-\theta_b^{(1)}
\end{array}
\eea when $k=0$ or
\bea\label{C_ij_case2n1}
&&|C_{12}|^2=\sin^2(\theta_b^{(1)}/2),\\[1ex]
\nonumber
&&|C_{13}|^2=\Biggl[\begin{array}{ccc}\cos^2\theta_b^{(1)}&
\mbox{at}&\theta_b^{(3)}=\theta_b^{(1)},\\
0&
\mbox{at}&\qquad\theta_b^{(3)}=\pi-\theta_b^{(1)},\end{array}\\
\nonumber
&&|C_{23}|^2=\Biggl[\begin{array}{ccc}\sin^2(\theta_b^{(1)}/2)&
\mbox{at}&\!\!\!\theta_b^{(3)}=\theta_b^{(1)},\\
\cos^2(\theta_b^{(1)}/2)&
\mbox{at}&\quad\theta_b^{(3)}=\pi-\theta_b^{(1)}
\end{array}
\eea when $k=1$.

It is seen that contrary to the case 1, this kind of the
measurement settings needs only a single polar angle
$\theta_b^{(1)}$.
 \vspace{-7mm}
\subsection{The case $\sin\theta_b^{(1)}\ne 0$, $\sin\theta_b^{(2)}\ne
0$, $\phi_b^{(i)}-\phi_b^{(j)}\ne\pi k$. \label{case3}}
\vspace{-4mm} Having denoted
$\Delta_{ij}=\phi_b^{(i)}-\phi_b^{(j)}$, one can express
$\theta_b^{(2)}$ and $\theta_b^{(3)}$ via the single polar angle
$\theta_b^{(1)}$ as follows:
\bea\label{thet2_thet3} \nonumber &&
\sin\theta_b^{(2)}=\frac{\sin\Delta_{13}}{\sin(\Delta_{12}-\Delta_{13})}\sin\theta_b^{(1)},
\\
&&
\sin\theta_b^{(3)}=-\frac{\sin\Delta_{12}}{\sin(\Delta_{12}-\Delta_{13})}\sin\theta_b^{(1)}.
\eea
 Taking into account the restriction (\ref{0pi}), it is straightforward to verify that
Eqs.~(\ref{thet2_thet3}) are satisfied at simultaneous realization
of the following conditions:
\begin{widetext}\bea
\label{Acond_case3} \left\{
\begin{array}{c} 0<\Delta_{13}<\pi,\\
\pi<\Delta_{12}< 2\pi,\\
0<\Delta_{12}-\Delta_{13}<\pi,\\
\Delta_{13}+\arcsin[\sin\Delta_{13}\sin\theta_b^{(1)}]\le\Delta_{12}\le\pi+\Delta_{13}-\arcsin[\sin\Delta_{13}
\sin\theta_b^{(1)}],\\
\Delta_{12}-\pi-\arcsin[\sin\Delta_{12}\sin\theta_b^{(1)}]\le\Delta_{13}\le\Delta_{12}+\arcsin[\sin\Delta_{12}
\sin\theta_b^{(1)}].
\end{array}
\right. \eea \end{widetext} It is seen from
Eq.~(\ref{thet2_thet3}) that the case 3 implies the
three-parameter measurement problem. However, like in the case 1,
there exists a restriction on the possible values of the
parameters $\Delta_{12}$, $\Delta_{23}$ and $\theta_b^{(1)}$ (see
Eq.~ (\ref{Acond_case3}) that, once again, yields a ``window'' for
the allowed Euler angles.

\section{Measurement settings that ensure
condition $\omega_m^{(ext)}=1/3$.}

\renewcommand{\theequation}{B.\arabic{equation}}
\vspace{-2mm} In the \underline{\textbf{case 1}} this condition is
realized at $\theta_a^{(m)}=\pi/2$ and yields
$|C_{12}|^2=|C_{13}|^2=|C_{23}|^2$. The last equality at $k=l=1$
due to (\ref{mn})-(\ref{Cijcase1}) gives us the system of
equations for three polar angles: \bea\label{Cequal1}
\left\{\begin{array}{c}
\theta_b^{(1)}+\theta_b^{(2)}+2 \theta_b^{(3)}=2\pi p,\\
2\theta_b^{(1)}-\theta_b^{(2)}+\theta_b^{(3)}=2\pi q,\\
-\theta_b^{(1)}+2\theta_b^{(2)}+\theta_b^{(3)}=2\pi r,
\end{array}\right. \eea
where $\{p,q,r\}=\{0,1\}$. The system of equations (\ref{Cequal1})
is compatible only at $r+q=p$. To satisfy this condition, we can
choose $q=0$, $r=1$ and $p=1$ that yields \bea\label{Cequal1-1}
\theta_b^{(1)}=\frac{2\pi}{3}-\theta_b^{(3)},\,\,\theta_b^{(2)}=\frac{4\pi}{3}-\theta_b^{(3)},\,\,
\frac{\pi}{3}<\theta_b^{(3)}<\frac{2\pi}{3} .\eea Another possible
choice $q=1$, $r=0$, $p=1$ gives us a similar result:
\bea\label{Cequal1-2}
\theta_b^{(1)}=\frac{4\pi}{3}-\theta_b^{(3)},\,\,\theta_b^{(2)}=\frac{2\pi}{3}-\theta_b^{(3)},\,\,
\frac{\pi}{3}<\theta_b^{(3)}<\frac{2\pi}{3} .\eea

If we select $l=0$, $k=1$, see Eqs.~(\ref{mn})-(\ref{Cijcase1}),
we obtain the system of equations \bea\label{Cequal2}
\left\{\begin{array}{c}
2\theta_b^{(1)}+\theta_b^{(2)}+\theta_b^{(3)}=2\pi p,\\
\theta_b^{(1)}+2\theta_b^{(2)}-\theta_b^{(3)}=2\pi q,\\
\theta_b^{(1)}-\theta_b^{(2)}+2\theta_b^{(3)}=2\pi r,
\end{array}\right. \eea
The system compatibility condition $r+q=p$ gives us two possible
realizations:
 \bea\label{Cequal2-1}
\theta_b^{(1)}=\frac{4\pi}{3}-\theta_b^{(3)},\,\,\theta_b^{(2)}=-\frac{2\pi}{3}+\theta_b^{(3)},\,\,
\frac{2\pi}{3}<\theta_b^{(3)}<\pi \eea at $q=0$, $r=1$, $p=1$, or
\bea\label{Cequal2-2}
\theta_b^{(1)}=\frac{2\pi}{3}-\theta_b^{(3)},\,\,\theta_b^{(2)}=\frac{2\pi}{3}+\theta_b^{(3)},\,\,
0<\theta_b^{(3)}<\frac{\pi}{3}\eea at $q=1$, $r=0$, $p=1$.

In the \underline{\textbf{case 2}} a simple analysis of
Eqs.~(\ref{C_ij_case2n0})-(\ref{C_ij_case2n1}) shows that
conditions $|C_{12}|^2=|C_{13}|^2=|C_{23}|^2$ leading to equation
$\omega_m^{(ext)}=1/3$ are satisfied at
\bea\label{CequalCase2}\nonumber &&\theta_b^{(1)}=2\pi/3 \qquad
\mbox{if}\quad
k=0,\\
&&\theta_b^{(1)}=\pi/3 \qquad\,\,\, \mbox{if}\quad k=1.\eea The
case $\theta_b^{(3)}=\pi-\theta_b^{(1)}$ has to be excluded from
consideration since it gives $\omega_m^{(ext)}=\{1/2,0,1/2\}$.
This result is unphysical (see Eq.~(\ref{rho-SApp})) because we
have postulated $a_i^{(m)}\ne 0$ for all the values of
$m=\{1,2,3\}$.

A detailed analysis of the \underline{\textbf{case 3}} shows that
the values (\ref{C_ijGen}) can never be equal to each other (see
also Eq.~(\ref{thet2_thet3})). Hence, the condition
$\omega_m^{(ext)}=1/3$ cannot be realized.

\end{document}